\documentclass[aps,prb,twocolumn,superscriptaddress,nofootinbib]{revtex4-2}
\usepackage{amsmath,amssymb,amsthm,bm}
\usepackage{graphicx}
\usepackage[colorlinks=true,allcolors=blue]{hyperref}

\newcommand{\ket}[1]{\lvert #1\rangle}
\newcommand{\bra}[1]{\langle #1\rvert}

\begin{document}
\title{From the Full Fermionic Hilbert Space to the BCS Paired Subspace}
\author{Guo-Jian Qiao}
\affiliation{Graduate School of China Academy of Engineering Physics, Beijing 100193, China}
\author{Xin Yue}
\affiliation{Beijing Computational Science Research Center, Beijing 100193, China}

\author{C. P. Sun}
\email{suncp@gscaep.ac.cn}
\affiliation{Graduate School of China Academy of Engineering Physics, Beijing 100193, China}

\begin{abstract}
Conventional Bardeen--Cooper--Schrieffer (BCS) and number-projected BCS variational treatments are formulated within a paired Hilbert subspace, where modes with opposite momenta and spins are either both empty or both occupied. Here we ask whether this paired subspace can itself be selected energetically from the full fermionic Hilbert space of the attractive reduced BCS model. Using conserved projectors that distinguish singly occupied states from empty and doubly occupied ones, we decompose the full Hilbert space into invariant sectors. A direct energy comparison between sectors shows that, at zero temperature, every ground state at fixed average particle
number lies in the paired subspace. The same holds at fixed even particle number, with the comparison preserving the total particle number, while at fixed odd particle
number one fermion remains unpaired, as required by parity. Within the resulting paired subspace, minimization over positive generalized-$P$ representations of pseudospin coherent states selects a common phase for the nonzero pair amplitudes and determines their magnitudes. BCS and number-projected BCS states arise as particular minimizers under the respective particle-number constraints.
\end{abstract}

\maketitle

\section{Introduction}
In the conventional Bardeen--Cooper--Schrieffer (BCS) variational
treatment, the trial space is restricted from the outset to states in
which the two modes $(\bm k\uparrow,-\bm k\downarrow)$ are either both
empty or both occupied. Specifically, the variational
state is 
\begin{equation}
\ket{\mathrm{BCS}}
=
\prod_{\bm k}
\left(
u_{\bm k}
+
v_{\bm k}e^{i\phi_{\bm k}}
\hat c_{\bm k\uparrow}^{\dagger}
\hat c_{-\bm k\downarrow}^{\dagger}
\right)\ket{0}.
\label{eq:introBCS}
\end{equation}
Minimization with respect to the phases $\phi_{\bm k}$ shows that
the attractive interaction favors a common phase for the nonzero pair
amplitudes, while subsequent variation of their magnitudes gives the
familiar BCS self-consistency equations \cite{BCS1957}.

A closely related variational treatment is formulated in the sector
of the Hilbert space with fixed particle number $N=2M$. These states are obtained by projecting BCS
states onto the $N$-particle sector and can equivalently be written
as condensates of $M$ pairs:
\begin{equation}
\ket{\Psi_M}
\propto
\left(
\sum_{\bm k}
g_{\bm k}
\hat c_{\bm k\uparrow}^{\dagger}
\hat c_{-\bm k\downarrow}^{\dagger}
\right)^M
\ket{0}.
\label{eq:introPBCS}
\end{equation}
Within this variational space, Bayman showed that the BCS equations
can be recovered through a saddle-point approximation
\cite{Bayman1960}, and the BCS wave functions with a fixed particle
number have subsequently been studied
\cite{Dietrich1964,IrkhinKatsnelson1984,Leggett2006,Dukelsky2016}. 

Although the BCS state in Eq.~\eqref{eq:introBCS} generally contains
components with different particle numbers and the number-projected
BCS state in Eq.~\eqref{eq:introPBCS} has exactly $N$ particles, both
are constructed within the paired subspace
\begin{equation}
\mathcal H_{\mathrm{pair}}
=
\bigotimes_{\bm k}
\operatorname{span}
\left\{
\ket{0}_{\bm k},
\ket{P}_{\bm k}
\right\},
\label{eq:intro_pairspace}
\end{equation}
where
$\ket{P}_{\bm k}
=
\hat c_{\bm k\uparrow}^{\dagger}
\hat c_{-\bm k\downarrow}^{\dagger}
\ket{0}_{\bm k}$.
The singly occupied states
$\ket{\uparrow}_{\bm k}
=
\hat c_{\bm k\uparrow}^{\dagger}\ket{0}_{\bm k}$
and
$\ket{\downarrow}_{\bm k}
=
\hat c_{-\bm k\downarrow}^{\dagger}\ket{0}_{\bm k}$ lie outside this subspace and are therefore excluded before energy minimization.

This observation motivates a more precise question than whether pairing
occurs in the BCS model: can the restriction to
$\mathcal H_{\mathrm{pair}}$ itself be derived from energy minimization
over the full fermionic Hilbert space, rather than imposed as part of the
variational ansatz? A second question then arises within the selected
paired subspace: whether the common phase of pair amplitudes in the BCS
state follows from the same energetic minimization when the variational
class is enlarged beyond a single product coherent state?

\begin{figure}[t]
\centering
\includegraphics[width=0.90\columnwidth]{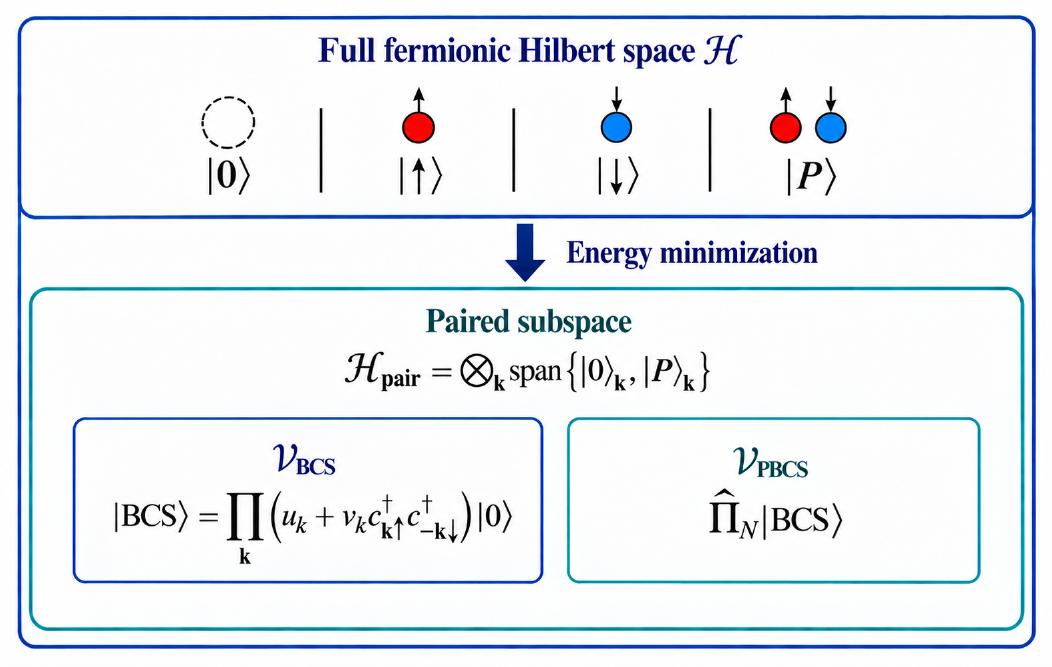}
\caption{The full Hilbert space $\mathcal{H}$ contains empty, singly occupied, and doubly occupied momentum modes. Energy minimization in the full Hilbert space excludes singly occupied modes and selects the paired subspace $\mathcal{H}_{\mathrm{pair}}$. Within this subspace, BCS states and number-projected BCS states form the two variational classes $\mathcal V_{\mathrm{BCS}}$ and $\mathcal V_{\mathrm{PBCS}}$, respectively. Energy minimization over positive generalized-$P$ representations selects a common phase for the nonzero pair amplitudes even beyond
these two variational classes.}
\label{fig:BCS_PiN}
\end{figure}

In this paper, we first consider the reduced BCS Hamiltonian on the full fermionic Hilbert space, without restricting the variational space to BCS product states or number-projected BCS states. Using the conserved projectors, which distinguish singly occupied from empty or doubly occupied momentum channels, the Hilbert space is decomposed into invariant sectors. Direct comparison of the energies of these sectors shows that, at zero temperature, every ground state under either fixed average particle number or fixed even particle number contains no singly occupied momentum modes (see Fig. \ref{fig:BCS_PiN}). For odd particle number, only the single unpaired fermion required by number parity remains. Thus, the fully paired structure of the superconducting ground state is not an assumption built into the BCS ansatz, but follows directly from energy minimization in the full fermionic Hilbert space.

We next use pseudospin coherent states and their number-projected counterparts to construct generalized-$P$ representations at fixed average particle number and fixed particle number. In both cases, energy minimization over positive generalized-$P$ representations selects a common phase for the nonzero pair amplitudes at different
momenta, which we term the phase coherence of Cooper
pairs. The BCS state and the number-projected BCS state arise as particular minimum-energy solutions under the respective particle-number constraints. At fixed particle number, projection makes the common phase physically irrelevant, while the relative
phases remain equal. Phase coherence thus follows from minimization over a class broader than either BCS variational class.

\section{Absence of single occupancy in the ground state}
\label{sec:selection}

We consider the reduced BCS Hamiltonian \cite{BCS1957,Richardson1963}
\begin{equation}
	\hat H=
	\sum_{\bm{k},\sigma}
	\epsilon_{\bm{k}}
	\hat c_{\bm{k}\sigma}^{\dagger}\hat c_{\bm{k}\sigma}
	-\frac{g}{V}
	\sum_{\bm{k},\bm{k}'}
	\hat c_{\bm{k}\uparrow}^{\dagger}
	\hat c_{-\bm{k}\downarrow}^{\dagger}
	\hat c_{-\bm{k}'\downarrow}
	\hat c_{\bm{k}'\uparrow},
	\label{eq:H}
\end{equation}
where $g>0$ is the attractive interaction strength and $V$ is the system volume. The Hamiltonian conserves the total particle number,
$[\hat H,\hat N]=0$, with
$\hat N=\sum_{\bm{k},\sigma}
\hat c_{\bm{k}\sigma}^{\dagger}\hat c_{\bm{k}\sigma}$. In addition to $\hat N$, the Hamiltonian possesses a set of conserved
projectors associated with each pair of modes
$(\bm{k}\uparrow,-\bm{k}\downarrow)$, defined as \cite{Richardson1964,Delft2001}
\begin{equation}
	\hat Q_{\bm{k}}
	=
	\left(
	\hat n_{\bm{k}\uparrow}
	-
	\hat n_{-\bm{k}\downarrow}
	\right)^2,
	\label{eq:Qk}
\end{equation}
where $\hat n_{\bm{k}\sigma}=\hat c_{\bm{k}\sigma}^{\dagger}\hat c_{\bm{k}\sigma}$. These operators satisfy
\begin{equation}
	\hat Q_{\bm{k}}^2=\hat Q_{\bm{k}},
	\qquad
	[\hat H,\hat Q_{\bm{k}}]=0,
	\qquad
	[\hat Q_{\bm{k}},\hat Q_{\bm{k}'}]=0.
	\label{eq:Qcomm}
\end{equation}
Since $\hat Q_{\bm{k}}$ is a projector, its eigenvalues are $q_{\bm{k}}=0,1$. Indeed, the two-mode Hilbert space associated with
$(\bm{k}\uparrow,-\bm{k}\downarrow)$ is spanned by
$\ket{0}_{\bm{k}}$,
the two singly occupied states
$\ket{\uparrow}_{\bm{k}}$ and
$\ket{\downarrow}_{\bm{k}}$,
and the paired state
$\ket{P}_{\bm{k}}$. The empty and doubly occupied states,
$\ket{0}_{\bm{k}}$ and $\ket{P}_{\bm{k}}$, have $q_{\bm{k}}=0$,
whereas the singly occupied states,
$\ket{\uparrow}_{\bm{k}}$ and $\ket{\downarrow}_{\bm{k}}$, have
$q_{\bm{k}}=1$.

Because the projectors $\hat{Q}_{\bm{k}}$ mutually commute and are conserved by
$\hat H$, the full Hilbert space decomposes into a direct sum of invariant
sectors, $\mathcal H = \bigoplus_{\{q_{\bm{k}}\}} \mathcal H_{\{q_{\bm{k}}\}}$, where each sector is characterized by a fixed set of eigenvalues $\{q_{\bm{k}}\}$. The ground state energy can therefore be determined by comparing the lowest energies of the different invariant sectors. We now show that, at zero temperature, every ground state, either at fixed average particle number or at fixed particle number $N=2M$,
belongs to the fully paired subspace
\begin{equation}
	\mathcal H_{\mathrm{pair}}
	=
	\bigotimes_{\bm{k}}
	\operatorname{span}
	\left\{
	\ket{0}_{\bm{k}},
	\ket{P}_{\bm{k}}
	\right\}.
	\label{eq:pairedspace}
\end{equation}
Equivalently, the ground state contains no singly occupied momentum modes.

\subsection{Fixed average particle number}
Under the constraint of a fixed average particle number, the 
grand-canonical energy at zero temperature is
\begin{equation}
\mathcal F=\langle \hat H \rangle-\mu\langle\hat N\rangle ,
	\label{eq:Omega}
\end{equation}
where $\mu$ is the chemical potential. Suppose that a state contains a singly occupied mode labeled by
$(\bm q,\sigma)$. Define
\begin{equation}
	\hat{d}_{\bm q\sigma}^{\dagger}
	=
	\begin{cases}
		\hat{c}_{\bm q\uparrow}^{\dagger},
		& \sigma=\uparrow,\\[1mm]
		\hat{c}_{-\bm q\downarrow}^{\dagger},
		& \sigma=\downarrow .
	\end{cases}
	\label{eq:singlecreator}
\end{equation}
A normalized state with this single occupancy can be written as
\begin{equation}
	\ket{\Psi_{1}}
	=
	\hat{d}_{\bm q\sigma}^{\dagger}\ket{\Phi},
	\qquad
	\langle\Phi|\Phi\rangle=1,
	\label{eq:blockedgeneralstate}
\end{equation}
where both modes $\bm q\uparrow$ and $-\bm q\downarrow$ are empty in
$\ket{\Phi}$, while the occupations of all other modes are unrestricted
and may, in particular, include single occupancies.

Let $\mathcal F_{\bar{\bm q}}$ denote the contribution to the
grand-canonical energy from all modes except
$(\bm q\uparrow,-\bm q\downarrow)$. The singly occupied state then has
\begin{equation}
	\mathcal F_{1}
	=
	\mathcal F_{\bar{\bm q}}
	+
	\xi_{\bm q},
	\label{eq:Omega1}
\end{equation}
where $\xi_{\bm q}=\epsilon_{\bm q}-\mu$.
We compare this state with the corresponding empty and doubly occupied
states,
\begin{equation}
	\ket{\Psi_{0}}
	=
	\ket{\Phi},
	\qquad
	\ket{\Psi_{2}}
	=
	\hat c_{\bm q\uparrow}^{\dagger}
	\hat c_{-\bm q\downarrow}^{\dagger}
	\ket{\Phi}.
	\label{eq:trialstates}
\end{equation}
whose grand-canonical energies are
\begin{equation}
	\mathcal F_{0}
	=
	\mathcal F_{\bar{\bm q}},
	\qquad
	\mathcal F_{2}
	=
	\mathcal F_{\bar{\bm q}}
	+
	2\xi_{\bm q}
	-
	\frac{g}{V}.
	\label{eq:trialenergies}
\end{equation}
Comparing these expressions with the grand-canonical energy of the singly occupied state gives
\begin{equation}
	\mathcal F_{0}-\mathcal F_{1}
	=
	-\xi_{\bm q},
	\qquad
	\mathcal F_{2}-\mathcal F_{1}
	=
	\xi_{\bm q}
	-
	\frac{g}{V}.
	\label{eq:energydifferences}
\end{equation}
Since $g>0$, the two differences cannot both be nonnegative, and hence
\begin{equation}
	\min\{\mathcal F_{0},\mathcal F_{2}\}
	<
	\mathcal F_{1}.
	\label{eq:lowerenergy}
\end{equation}

Equation~\eqref{eq:lowerenergy} excludes every sector containing a
singly occupied mode from the ground state. Since $\bm q$ is arbitrary,
the ground state must satisfy $q_{\bm k}=0$ for all $\bm k$ and therefore
belongs to the fully paired subspace $\mathcal F_{\mathrm{pair}}$ defined
in Eq.~\eqref{eq:pairedspace}. This proves the fixed-average-particle-number case of the result stated above.

\subsection{Fixed particle number}
We next consider a fixed particle number $N$. Since $[\hat H,\hat N]=0$,
the Hilbert space at fixed $N$ is
\begin{equation}
	\mathcal H_N
	=
	\left\{
	\ket{\Psi}\in\mathcal H
	\;\middle|\;
	\hat N\ket{\Psi}=N\ket{\Psi}
	\right\}.
	\label{eq:fixedNspace}
\end{equation}
Moreover, since $[\hat N,\hat{Q}_{\bm k}]=0$, the decomposition associated
with the conserved projectors $Q_{\bm k}$ remains valid within
$\mathcal H_N$:
\begin{equation}
	\mathcal H_N
	=
	\bigoplus_{\{q_{\bm k}\}\in\mathcal C_N}
	\mathcal H_{N,\{q_{\bm k}\}},
	\label{eq:fixedN_decomposition}
\end{equation}
where $\mathcal C_N$ denotes the set of eigenvalue configurations
$\{q_{\bm k}\}$ compatible with the fixed particle number $N$.

For a given configuration, let
$b=\sum_{\bm k}q_{\bm k}$ denote the number of singly occupied modes.
Since $q_{\bm k}=0$ corresponds to either zero or double occupancy,
the total particle number satisfies
\begin{equation}
	N=b+2m,
	\qquad
	m=0,1,2,\ldots,
	\label{eq:Nbm}
\end{equation}
where $m$ is the number of doubly occupied modes. In particular, for
an even particle number $N=2M$, $b$ must also be even. Hence any sector
containing single occupancy necessarily contains at least two singly
occupied modes. Consider two of them, labeled by $\bm q$ and $\bm p$. Using
$\hat{d}_{\bm q\sigma}^{\dagger}$ defined in
Eq.~\eqref{eq:singlecreator}, a normalized state in this sector can be
written as
\begin{equation}
	\ket{\Psi_{11}}
	=
	\hat{d}_{\bm q\sigma}^{\dagger}
	\hat{d}_{\bm p\sigma'}^{\dagger}
	\ket{\Phi},
	\qquad
	\langle\Phi|\Phi\rangle=1,
	\label{eq:Psi11_fixedN}
\end{equation}
where $\ket{\Phi}$ contains the remaining $N-2$ particles and the
modes associated with $\bm q$ and $\bm p$ are empty. Apart from this
restriction, $\ket{\Phi}$ is arbitrary and may contain other single
occupancies.

Let $\mathcal F_{\overline{\bm q,\bm p}}$ denote the contribution to the energy
from all modes other than those associated with $\bm q$ and $\bm p$.
The energy of $\ket{\Psi_{11}}$ is
\begin{equation}
	\mathcal F_{11}
	=
	\mathcal F_{\overline{\bm q,\bm p}}
	+
	\epsilon_{\bm q}
	+
	\epsilon_{\bm p}.
	\label{eq:E11_fixedN}
\end{equation}
We now redistribute the same two fermions while keeping the total
particle number unchanged. Starting from the same $\ket{\Phi}$,
consider
\begin{equation}
	\ket{\Psi_{20}}
	=
	\hat{c}_{\bm q\uparrow}^{\dagger}
	\hat{c}_{-\bm q\downarrow}^{\dagger}
	\ket{\Phi},
	\qquad
	\ket{\Psi_{02}}
	=
	\hat{c}_{\bm p\uparrow}^{\dagger}
	\hat{c}_{-\bm p\downarrow}^{\dagger}
	\ket{\Phi}.
	\label{eq:Psi20_Psi02}
\end{equation}
Their energies are
\begin{equation}
	\mathcal F_{20}
	=
	\mathcal F_{\overline{\bm q,\bm p}}
	+
	2\epsilon_{\bm q}
	-
	\frac{g}{V},
	\qquad
	\mathcal F_{02}
	=
	\mathcal F_{\overline{\bm q,\bm p}}
	+
	2\epsilon_{\bm p}
	-
	\frac{g}{V}.
	\label{eq:E20_E02_fixedN}
\end{equation}
Comparing these energies with $F_{11}$ gives
\begin{equation}
	\left(\mathcal F_{20}-\mathcal F_{11}\right)
	+
	\left(\mathcal F_{02}-\mathcal F_{11}\right)
	=
	-\frac{2g}{V}
	<0.
	\label{eq:fixedN_energy_sum}
\end{equation}
Hence $\mathcal F_{20}$ and $\mathcal F_{02}$ cannot both be greater than or equal to
$\mathcal F_{11}$; at least one of the two configurations has strictly lower
energy.

Thus, two single occupancies can always be removed at fixed particle
number with a strict reduction in energy. Repeating this comparison
eliminates single occupancies pairwise. Since $b$ is even for $N=2M$,
the process terminates at $b=0$. The fixed-$N$ ground state therefore
belongs to
\begin{equation}
\mathcal H_{2M}^{\mathrm{pair}}=\operatorname{span}
	\left\{
	\prod_{\bm k\in S}
	\hat{c}_{\bm k\uparrow}^{\dagger}
	\hat{c}_{-\bm k\downarrow}^{\dagger}
	\ket{0}
	\;\middle|\;
	|S|=M
	\right\}.
	\label{eq:fixedN_pair_sector}
\end{equation}
This proves the fixed-particle-number case of the result stated above. For completeness, when $N=2M+1$ is odd, Eq.~\eqref{eq:Nbm} requires $b$ to be odd. The same pairwise argument eliminates single occupancies until $b=1$, leaving one unpaired fermion together with $M$ doubly occupied modes. This even--odd distinction agrees with the familiar blocking picture of the
reduced BCS model~\cite{Delft2001}. When all $\epsilon_{\bm k}$ are equal, our argument also recovers the result that the even-$N$ ground state has no singly occupied modes~\cite{Richardson1964}.

These results show that, for the attractive Hamiltonian considered here, pairing is not imposed as a variational assumption. Rather, it follows from energy minimization in the full fermionic Hilbert space: at fixed average particle number and at fixed even particle number, the ground state contains no singly occupied modes and is confined to the paired subspace. For odd particle number, the same argument leaves only the single unpaired fermion required by particle-number parity. Thus, the paired structure of the ground state emerges as an energetic consequence of the attractive interaction itself.

\section{Energetics and coherence of paired states}
\label{sec:coherent}

In the previous section, we showed that energy minimization excludes single occupancy from the ground state at fixed average particle number and at fixed even particle number. We now examine how the amplitudes and relative phases of Cooper pairs are determined within the paired subspace.

\subsection{Pseudospin coherent states and the $P$ representation}

Within the paired subspace, we introduce the Anderson pseudospin
operators  \cite{Anderson1958}
$$\hat{S}_{\bm k}^{+}
	=
	(\hat{S}_{\bm k}^{-})^{\dagger}
	=
	\hat{c}_{\bm k\uparrow}^{\dagger}
	\hat{c}_{-\bm k\downarrow}^{\dagger},
	\quad
	\hat{S}_{\bm k}^{z}
	=
	\frac{1}{2}
	\left(
	\hat{n}_{\bm k\uparrow}
	+
	\hat{n}_{-\bm k\downarrow}
	-1
	\right).$$
which satisfy the $SU(2)$ algebra
\begin{equation}
	[\hat{S}_{\bm k}^{+},\hat{S}_{\bm k}^{-}]
	=
	2\hat{S}_{\bm k}^{z},
	\qquad
	[\hat{S}_{\bm k}^{z},\hat{S}_{\bm k}^{\pm}]
	=
	\pm \hat{S}_{\bm k}^{\pm}.
	\label{eq:SU2algebra}
\end{equation}
The empty state $\ket{0}_{\bm k}$ and the doubly occupied state
$\ket{P}_{\bm k}$ therefore form a spin-$1/2$ representation of $SU(2)$. The associated pseudospin coherent state is
\begin{equation}
	\ket{\Omega_{\bm k}}
	=
	\cos\frac{\theta_{\bm k}}{2}\ket{0}_{\bm k}
	+
	e^{i\phi_{\bm k}}
	\sin\frac{\theta_{\bm k}}{2}\ket{P}_{\bm k},
	\label{eq:localcoherent}
\end{equation}
where $0\leq\theta_{\bm k}\leq\pi$ and
$0\leq\phi_{\bm k}<2\pi$. It resolves the identity in the
two-dimensional paired space,
\begin{equation}
	\mathbf 1_{\bm k}
	=
	\int
	\frac{\sin\theta_{\bm k}\,
		d\theta_{\bm k}\,d\phi_{\bm k}}{2\pi}
	\ket{\Omega_{\bm k}}
	\bra{\Omega_{\bm k}}.
	\label{eq:localidentity}
\end{equation}

For the full paired Hilbert space, we introduce the product coherent
state and the corresponding measure,
\begin{equation}
	\ket{\bm\Omega}
	=
	\bigotimes_{\bm k}\ket{\Omega_{\bm k}},
	\qquad
	d\mu(\bm\Omega)
	=
	\prod_{\bm k}
	\frac{\sin\theta_{\bm k}\,
		d\theta_{\bm k}\,d\phi_{\bm k}}{2\pi}.
	\label{eq:globalcoherent}
\end{equation}
Taking the tensor product of Eq.~\eqref{eq:localidentity} over all
$\bm k$ gives
\begin{equation}
	\mathbf 1_{\mathrm{pair}}
	=
	\int d\mu(\bm\Omega)\,
	\ket{\bm\Omega}\bra{\bm\Omega}.
	\label{eq:globalidentity}
\end{equation}
Thus, the product pseudospin coherent states form an overcomplete set spanning the entire paired Hilbert space. The completeness relation in Eq.~\eqref{eq:globalidentity} allows a density operator in the paired Hilbert space to be represented in the
generalized diagonal form~\cite{Agarwal1981,Giraud2008},
\begin{equation}
	\rho
	=
	\int d\mu(\bm\Omega)\,
	P(\bm\Omega)
	\ket{\bm\Omega}\bra{\bm\Omega},
	\label{eq:Prepresentation}
\end{equation}
where $\int d\mu(\bm\Omega)\,P(\bm\Omega)=1$. In general, $P(\bm\Omega)$ is a generalized distribution and need not be positive.

The same construction can be carried out under a strict particle-number constraint. For a fixed even particle number $N=2M$, the ground state is confined to the fully paired sector established above. The particle-number
projector is
\begin{equation}
	\hat{\Pi}_N
	=
	\frac{1}{2\pi}
	\int_{0}^{2\pi}
	d\chi\,
	e^{i\chi(\hat N-N)} .
	\label{eq:number_projector}
\end{equation}
Projecting the coherent states introduced above gives
\begin{equation}
	\ket{\bm\Omega;N}
	=
	\frac{\hat{\Pi}_N\ket{\bm\Omega}}
	{\sqrt{\mathcal N_N(\bm\Omega)}},
	\qquad
	\mathcal N_N(\bm\Omega)
	=
	\bra{\bm\Omega}\hat{\Pi}_N\ket{\bm\Omega}.
	\label{eq:projected_coherent}
\end{equation}
Projecting Eq.~\eqref{eq:globalidentity} onto the $N$-particle sector
then yields
\begin{equation}
	\mathbf {I}_N^{\rm pair}
	=
	\int d\nu_N(\bm\Omega)\,
	\ket{\bm\Omega;N}
	\bra{\bm\Omega;N},
	\label{eq:projected_identity}
\end{equation}
where
$d\nu_N(\bm\Omega)
=d\mu(\bm\Omega)\mathcal N_N(\bm\Omega)$.
The projected coherent states are therefore complete in the paired
subspace with fixed particle number $N$. A density operator in this
subspace admits a generalized-$P$ representation,
\begin{equation}
	\rho_N
	=
	\int d\nu_N(\bm\Omega)\,
	P_N(\bm\Omega)
	\ket{\bm\Omega;N}\bra{\bm\Omega;N},
	\label{eq:PNrepresentation}
\end{equation}
with $\int d\nu_N(\bm\Omega)\,P_N(\bm\Omega)=1$.

Equations~\eqref{eq:Prepresentation} and \eqref{eq:PNrepresentation} establish the generalized $P$ representations for the paired Hilbert space and its fixed-$N$ subspace, respectively. In principle, the variational problem may be formulated over all generalized distributions $P(\bm\Omega)$ or $P_N(\bm\Omega)$ that represent physical density operators. In the following, the optimization is restricted to generalized distributions that are positive in the distributional sense \cite{Giraud2008,Yao_2020}. Together with normalization, this class includes both regular probability densities and singular positive distributions such as delta functions.

\subsection{Phase coherence and the BCS state}

For a fixed average particle number, the zero-temperature
grand-canonical energy is obtained by
\begin{equation}
	\mathcal F [P]
	=
	\int d\mu(\bm\Omega)\,
	P(\bm\Omega)\,
	\mathcal E(\bm\Omega),
	\label{eq:Pfunctional}
\end{equation}
where
\begin{align}
	\mathcal E(\bm\Omega)
	&=
	\sum_{\bm k}
	\xi_{\bm k}
	\left(1-\cos\theta_{\bm k}\right)
	-
	\frac{g}{4V}
	\left|
	\sum_{\bm k}
	e^{i\phi_{\bm k}}
	\sin\theta_{\bm k}
	\right|^2
	\nonumber\\
	&-
	\frac{g}{4V}
	\sum_{\bm k}
	\left(1-\cos\theta_{\bm k}\right)^2.
	\label{eq:coherentenergy}
\end{align}
Restricting to the positive generalized-$P$ class introduced above, $\mathcal F[P]$ is a normalized weighted average of the coherent-state energy $\mathcal E(\bm\Omega)$. Defining $\mathcal E_{\min}=\min_{\bm\Omega}\mathcal E(\bm\Omega)$, positivity and normalization of $P$ give
\begin{equation}
	\mathcal F [P]-\mathcal E_{\min}
	=
	\int d\mu(\bm\Omega)\,P(\bm\Omega)
	\left[\mathcal E(\bm\Omega)-\mathcal E_{\min}\right]
	\geq0.
	\label{eq:Pminimum}
\end{equation}
Hence $\mathcal F [P]\geq\mathcal E_{\min}$, with equality only when
$P(\bm\Omega)$ is supported on configurations satisfying
$\mathcal E(\bm\Omega)=\mathcal E_{\min}$. The minimization over
positive generalized distributions therefore reduces to finding the
minimum-energy configurations $\bm\Omega^{*}$ of
$\mathcal E(\bm\Omega)$.

Minimization with respect to the phases $\phi_{\bm k}$ gives
\begin{equation}
	\frac{g}{2V}
	\sin\theta_{\bm k}
	\sum_{\bm p}
	\sin\theta_{\bm p}
	\sin(\phi_{\bm k}-\phi_{\bm p})
	=
	0.
	\label{eq:phase_equation}
\end{equation}
For $g>0$, the minimum-energy solution among these stationary
configurations has a common phase for all nonvanishing pair amplitudes \cite{Kouzoudis2009,qiao2026,yi2026},
\begin{equation}
	\phi_{\bm k}^{*}
	=
	\phi
	\pmod{2\pi}.
	\label{eq:phase_locking}
\end{equation}
Thus Cooper-pair phase coherence follows directly from energy minimization. Minimization with respect to the amplitudes $\theta_{\bm k}$ then gives
\begin{equation}
	\left[
	\xi_{\bm k}
	-
	\frac{g}{2V}
	\left(1-\cos\theta_{\bm k}^{*}\right)
	\right]
	\sin\theta_{\bm k}^{*}
	-
	\Delta
	\cos\theta_{\bm k}^{*}
	=
	0,
	\label{eq:amplitude_equation}
\end{equation}
where
\begin{equation}
	\Delta
	=
	\frac{g}{2V}
	\sum_{\bm k}
	\sin\theta_{\bm k}^{*}.
	\label{eq:gap_definition}
\end{equation}
Accordingly, the phase coherence and amplitudes of Cooper pairs are determined by energy minimization within the positive generalized-$P$ class. The minimum-energy distributions are supported on the corresponding configurations $\bm\Omega^{*}$. Two particular choices are especially simple.

First, the distribution may be concentrated at a single minimum
$\bm\Omega_0^{*}$,
\begin{equation}
	P_{\rm loc}(\bm\Omega)
	=
	\delta_{\mu}(\bm\Omega,\bm\Omega_0^{*}),
	\label{eq:localizedP}
\end{equation}
where $\delta_{\mu}$ denotes the delta distribution with respect to
$d\mu(\bm\Omega)$. The corresponding density operator is $\rho_{\rm loc}
	=
	\ket{\bm\Omega_0^{*}}
	\bra{\bm\Omega_0^{*}}$. Using the phase-locking condition, the state takes the form
\begin{equation}
	\ket{\mathrm{BCS}(\phi)}
	\equiv
	\ket{\bm\Omega_0^{*}}
	=
	\prod_{\bm k}
	\left[
	\cos\frac{\theta_{\bm k}^{*}}{2}
	+
	e^{i\phi}
	\sin\frac{\theta_{\bm k}^{*}}{2}
	\hat{S}_{\bm k}^{+}
	\right]\ket{0}.
\label{eq:BCSvariational}
\end{equation}
Equation~\eqref{eq:BCSvariational} is precisely the conventional BCS
wave function. Since $\bm\Omega_0^{*}$ minimizes
$\mathcal E(\bm\Omega)$, the BCS state is therefore a particular
minimum-energy solution within the positive generalized-$P$ class,
with its amplitudes determined by
Eqs.~\eqref{eq:amplitude_equation} and
\eqref{eq:gap_definition}.

Second, Eq.~\eqref{eq:phase_locking} leaves the common phase arbitrary.
The minimum-energy configurations therefore form a continuously
degenerate family, $\bm\Omega^{*}(\phi)
	=
	\{\theta_{\bm k}^{*},\phi\}$ with $0\leq\phi<2\pi$. Any positive normalized distribution supported on this family is also
a minimum-energy solution,
\begin{equation}
	\rho
	=
	\int_{0}^{2\pi}
	d\phi\,P(\phi)\,
	\ket{\mathrm{BCS}(\phi)}
	\bra{\mathrm{BCS}(\phi)}.
	\label{eq:phase_distribution}
\end{equation}
Here $P(\phi)$ is normalized as
$\int_{0}^{2\pi}d\phi\,P(\phi)=1$.
All states in this family have the same amplitudes
$\theta_{\bm k}^{*}$ and the same energy, differing only in the common
phase $\phi$. They therefore constitute a continuously degenerate
minimum-energy manifold within the positive generalized-$P$ class.

\subsection{Phase coherence and the number-projected BCS state}
For a fixed even particle number $N=2M$, we also restrict $P_N$
to positive generalized distributions, as in the preceding analysis. The
corresponding energy functional is
\begin{equation}
	\mathcal F_N[P_N]
	=
	\int d\nu_N(\bm\Omega)\,
	P_N(\bm\Omega)\,
	\mathcal E_N(\bm\Omega),
	\label{eq:fixedN_energy}
\end{equation}
with
$\mathcal E_N(\bm\Omega)
=\bra{\bm\Omega;N}\hat H\ket{\bm\Omega;N}$.
If
$\mathcal E_{N,\min}
=\min_{\bm\Omega}\mathcal E_N(\bm\Omega)$,
positivity and normalization of $P_N$ imply
\begin{equation}
	\mathcal F_N[P_N]-\mathcal E_{N,\min}
	=
	\int d\nu_N\,P_N
	\left(
	\mathcal E_N-\mathcal E_{N,\min}
	\right)
	\geq0.
	\label{eq:fixedN_minimum}
\end{equation}
Thus, within this variational class, energy minimization reduces to finding the projected coherent states that minimize
$\mathcal E_N(\bm\Omega)$.

To evaluate this energy, define
$z_{\bm k}=e^{i\phi_{\bm k}}\tan(\theta_{\bm k}/2)$ and
$x_{\bm k}=|z_{\bm k}|^2$. Let
$e_m=\sum_{\bm k_1<\cdots<\bm k_m}
x_{\bm k_1}\cdots x_{\bm k_m}$, with $e_0=1$, denote the
elementary symmetric polynomial of degree $m$ in
$\{x_{\bm k}\}$. The normalized projected state is \cite{Dietrich1964,Dukelsky2016}
$$\ket{\bm\Omega;N}
	=
	\frac{1}{\sqrt{e_M}}
	\sum_{\bm k_1<\cdots<\bm k_M}
	\left(
	\prod_{j=1}^{M}z_{\bm k_j}
	\right)
	\hat{S}_{\bm k_1}^{+}\cdots
	\hat{S}_{\bm k_M}^{+}\ket{0}.$$
We write $e_m^{(S)}$ for the same polynomial with the variables
indexed by $S$ excluded. Direct evaluation gives
\begin{equation}
	\mathcal E_N(\bm\Omega)
	=
	-\frac{gM}{V}
	+
	\frac{\mathcal A_M}{e_M},
	\label{eq:projected_energy}
\end{equation}
where
\begin{equation}
	\mathcal A_M
	=
	2\sum_{\bm k}
	\epsilon_{\bm k}x_{\bm k}
	e_{M-1}^{(\bm k)}-
	\frac{g}{V}
	\sum_{\bm k\neq\bm l}
	z_{\bm k}^{*}z_{\bm l}
	e_{M-1}^{(\bm k,\bm l)}.
	\label{eq:AM}
\end{equation}

Writing $z_{\bm k}=r_{\bm k}e^{i\phi_{\bm k}}$,
minimization with respect to $\phi_{\bm k}$ gives
\begin{equation}
	\sum_{\bm l\neq\bm k}
	r_{\bm k}r_{\bm l}
	e_{M-1}^{(\bm k,\bm l)}
	\sin(\phi_{\bm k}-\phi_{\bm l})
	=
	0.
	\label{eq:fixedN_phase_equation}
\end{equation}
For $g>0$, the minimum is attained when $\phi_{\bm k}^{*}=\phi\pmod{2\pi}$
for all coupled modes with nonzero amplitudes \cite{Dietrich1964}. Thus the phases
of the pair amplitudes are locked to be the same value.

On the phase-locked branch, the common phase may be removed, so that
$z_{\bm k}=r_{\bm k}\geq0$. For finite, nonzero amplitudes, using
$\partial e_M/\partial r_{\bm k}
=2r_{\bm k}e_{M-1}^{(\bm k)}$, minimization with respect to $r_{\bm k}$ gives
\begin{equation}
\frac{\partial\mathcal A_M}{\partial r_{\bm k}}
-
2r_{\bm k}e_{M-1}^{(\bm k)}
\left(
\mathcal E_N^{*}
+
\frac{gM}{V}
\right)
=
0,
\label{eq:fixedN_amplitude}
\end{equation}  
where $\mathcal E_N^{*}=\mathcal E_N(\bm\Omega^{*})$. Equations~\eqref{eq:fixedN_phase_equation} and
\eqref{eq:fixedN_amplitude} therefore show that relative Cooper-pair
phase coherence and the projected amplitude equations follow directly
from energy minimization at fixed particle number.

A particularly simple minimum-energy distribution is localized at one
projected minimum, $P_N^{\rm loc}(\bm\Omega)
	=
	\delta_{\nu_N}(\bm\Omega,\bm\Omega^{*})$.
It gives the pure state
\begin{equation}\rho_N^{\rm loc}
=\ket{\bm\Omega^{*};N}\bra{\bm\Omega^{*};N},
\end{equation} where
\begin{equation}
	\ket{\bm\Omega^{*};N}
	=
	\frac{1}
	{M!\sqrt{e_M(\bm x^{*})}}
	\left(
	\sum_{\bm k}
	r_{\bm k}^{*} \hat{S}_{\bm k}^{+}
	\right)^M
	\ket{0}.
	\label{eq:projected_BCS}
\end{equation}
Equation~\eqref{eq:projected_BCS} is the number-projected BCS state and
constitutes a particular minimum-energy solution within the positive
generalized-$P_N$ class. Finally, a common phase shift gives
\begin{equation}
	\ket{\bm\Omega^{*}(\phi);N}
	=
	e^{iM\phi}
	\ket{\bm\Omega^{*}(0);N}.
	\label{eq:fixedN_global_phase}
\end{equation}
Equation~\eqref{eq:fixedN_global_phase} shows explicitly that all values of
$\phi$ represent the same physical state at fixed particle number: the common phase contributes only an overall phase factor $e^{iM\phi}$ and is
not observable at fixed particle number.

We have shown that, at finite volume $V$, energy minimization within the positive generalized-$P$ and $P_N$ variational classes locks the relative phases of Cooper pairs and yields the respective amplitude equations, Eqs.~\eqref{eq:amplitude_equation} and \eqref{eq:fixedN_amplitude}. The BCS state and its number-projected counterpart arise from distributions concentrated at variational minima, rather than being specified in advance as the minimizing states. In the thermodynamic limit, the explicit $1/V$ term in Eq.~\eqref{eq:amplitude_equation} vanishes, giving the standard BCS gap equation at fixed average particle number. At fixed particle number, the saddle-point evaluation of number projection yields the same leading BCS variational energy in the bulk paired regime, and Eq.~\eqref{eq:fixedN_amplitude} consequently gives the same gap equation~\cite{Bayman1960,Braun1998}.

\section{Conclusion}

We have proved that energy minimization over the full fermionic
Hilbert space of the attractive reduced BCS model selects the paired
subspace. At finite volume $V$, a direct energy comparison excludes singly occupied modes from the ground state at fixed average particle number or fixed even particle number. For odd particle number, only the unpaired fermion required by parity remains.

Within the paired subspace, minimization over positive generalized-$P$ representations selects a common phase for nonzero Cooper-pair amplitudes and determines their magnitudes. The BCS state and its number-projected counterpart arise as minimizers within their respective variational classes. After number projection, the common phase becomes an overall phase of the state, while the relative phase coherence is retained. Together, these results provide an energetic justification for the paired Hilbert space adopted in BCS and number-projected BCS variational treatments.

\vspace{0.5em}
\noindent\textbf{Acknowledgments}---The authors thank Sheng‑Wen Li at BIT for helpful discussions on the variational approximation in \(P\)-representation. This study is supported by the National Natural Science Foundation of China (Grant Nos. 12688201 and 12547124), the China Postdoctoral Science Foundation (Grant Nos. 2025M784438 and GZC20261845), and the Science Challenge Project (Grant No. TZ2025017).

\vspace{0.5em}
\noindent\textbf{Data Availability}---No data were generated or analyzed in this theoretical study.

\enlargethispage{8\baselineskip}

\bibliographystyle{apsrev4-2}
\bibliography{Refs}

\end{document}